\documentstyle[12pt]{article}

\newcommand{\be}{\begin{equation}}
\newcommand{\ee}{\end{equation}}
\newcommand{\Dlt}{\Delta}
\newcommand{\dlt}{\delta}
\newcommand{\prt}{\partial}

\newcommand{\al}{\alpha}
\newcommand{\ra}{\rightarrow}

\newcommand{\lbd}{\lambda}
\newcommand{\om}{\omega}
\newcommand{\vp}{\varphi}

\begin{document}

\begin{center}
{\Large{\bf Fermi-Bose mapping for one-dimensional Bose gases} \\ [5mm]
V.I. Yukalov$^{1,2}$ and M.D. Girardeau$^3$} \\ [3mm]
{\it
$^1$Institut f\"ur Theoretische Physik, \\
Freie Universit\"at Berlin, Arnimallee 14, D-14195 Berlin, Germany \\ [2mm]

$^2$Bogolubov Laboratory of Theoretical Physics, \\
Joint Institute for Nuclear Research, Dubna 141980, Russia \\ [2mm]

$^3$Optical Science Center, \\
University of Arizona, Tucson, AZ 85721, USA}

\end{center}

\vskip 2cm

\begin{abstract}

One-dimensional Bose gases are considered, interacting either through the
hard-core potentials or through the contact delta potentials. Interest in
these gases gained momentum because of the recent experimental realization
of quasi-one-dimensional Bose gases in traps with tightly confined radial
motion, achieving the Tonks-Girardeau (TG) regime of strongly interacting 
atoms. For such gases the Fermi-Bose mapping of wavefunctions is applicable. 
The aim of the present communication is to give a brief survey of the 
problem and to demonstrate the generality of this mapping by emphasizing 
that: (i) It is valid for nonequilibrium wavefunctions, described by the
time-dependent Schr\"odinger equation, not merely for stationary 
wavefunctions. (ii) It gives the whole spectrum of all excited states, 
not merely the ground state. (iii) It applies to the Lieb-Liniger gas
with the contact interaction, not merely to the TG gas of
impenetrable bosons.

\end{abstract}

\vskip 2cm


\vskip 1cm


\newpage

\section{Introduction}

Physics of ultracold Bose gases is a rapidly developing field of research 
due to recent remarkable achievements in experiment and intensive theoretical
investigations (see reviews [1--4]). Among a number of important advancements
in this field, one of the major steps forward has been the realization of
quasi-one-dimensional Bose gases in elongated cylindrical traps and waveguides
by tightly confining the transverse atomic motion [5--19]. The TG regime of 
strongly interacting bosons has been reached in a one-dimensional 
optical-lattice trap [20]. The one-dimensional TG gas of bosons moving freely 
was also recently realized [21] for $^{87}$Rb atoms by trapping them with
a combination of two light traps. By changing the trap intensities it was 
possible to vary the ratio of the effective interaction to kinetic energy, 
achieving the TG regime, with this ratio reaching 5.5. The physical 
properties of low-dimensional Bose gases have been discussed in reviews 
[22,23]. However the important problem of the Fermi-Bose mapping in 
one-dimensional gases has not received the proper attention. It is the aim 
of the present short survey to compensate this deficiency by concentrating 
primarily on the Fermi-Bose mapping, first advanced over 40 years ago [24,25] 
for one-dimensional Bose gases. Particular emphasis will be placed on
demonstrating the generality of this mapping, whose applicability is
essentially wider than solely to the ground states of impenetrable bosons,
as it is often considered in literature, when the mapping is reduced to an
absolute-value relation. We shall stress that the mapping in its general 
form [24,25] is valid for the total spectrum of excited states, for the
time-dependent Schr\"odinger equation, and not only for impenetrable bosons, 
but also for the Lieb-Liniger gas with contact interactions of arbitrary
strength.

\section{TG gas}

The mapping theorem was formulated for a quantum one-dimensional system of
$N$ bosons at zero temperature [24,25]. First, only the stationary states were
considered. But, since the mapping procedure does not involve time, it was
later stressed that the same mapping also applies for the time dependent 
many-body Schr\"odinger equation, which was employed for treating temporal
interference properties of the one-dimensional hard-core Bose gas [26--31].
Therefore from the very beginning we may consider a time-dependent Hamiltonian
$\hat H\equiv \hat H(x_1,\ldots,x_N,t)$ where $x_i\in[0,L]$, with
$i=1,2,\ldots,N$. The Hamiltonian is written as a general expression
\be
\label{1}
\hat H =\hat K  + U + \frac{1}{2}\; \sum_{i\neq j}^N \; \Phi_{ij} \; ,
\ee
in which $\hat K$ is a kinetic-energy operator, having for nonrelativistic 
atoms of mass $m$ the form
\be
\label{2}
\hat K \equiv -\; \frac{\hbar^2}{2m} \; \sum_{i=1}^N 
\frac{\prt^2}{\prt x_i^2} \; ;
\ee
the term $U\equiv U(x_1,\ldots,x_N,t)$ describes any external potentials, 
such as confining potentials, generally allowing for the time dependence; and
$\Phi_{ij}\equiv \Phi(x_i-x_j)$ is a two-particle interaction potential. For
bosonic atoms, the wave function
\be
\label{3}
\psi_B \equiv\psi_B(x_1,\ldots,x_N,t)
\ee
is symmetric with respect to the permutations of any $x_i$ and $x_j$. The
wave function (3) is a solution to the time-dependent Schr\"odinger equation
\be
\label{4}
i\hbar\; \frac{\prt}{\prt t} \; \psi = \hat H \psi \; .
\ee
In the stationary case, this reduces to the eigenvalue problem
\be
\label{5}
\hat H\psi = E\psi \; .
\ee
The two-particle interaction is assumed to contain a hard core of diameter $a$,
which can be conveniently treated as a constraint on allowed wave functions
\be
\label{6}
\psi(x_1,\ldots,x_N,t) = 0 \qquad (|x_i - x_j| \leq a)
\ee
for $1\leq i< j\leq N$, rather than as an infinite contribution to the
interaction potential. Under constraint (6), the infinite hard-core potentials 
can be omitted from the Schr\"odinger equation, at the same time including all
other possible finite interactions into the term $U$. Then Hamiltonian (1)
can be contracted to the form
\be
\label{7}
\hat H =\hat K + U \; ,
\ee
while imposing constraint (6) on the solution of the Schr\"odinger equation.
The latter in a particular case of point-like impenetrable particles simplifies
to
\be
\label{8}
\psi(x_1,\ldots,x_N,t) = 0 \qquad (x_i=x_j) \; ,
\ee
where $1\leq i < j\leq N$.

The problem of a one-dimensional hard-core gas was raised by Tonks [32], who
considered the statistical mechanics of the classical high-temperature regime,
while Girardeau [24,25] gave the solution for the quantum problem. That is
why the one-dimensional system of impenetrable bosons is now commonly called
the Tonks-Girardeau gas.

The mapping theorem [24,25] can be formulated as follows. Let a wave function
$\psi_F\equiv\psi_F(x_1,\ldots,x_N,t)$ be a solution to the time-dependent
Schr\"odinger equation (4), with Hamiltonian (7), possessing the fermionic
antisymmetric property with respect to all permutations of any $x_i$ and $x_j$,
for $i\neq j$, and satisfying constraint (6). Introduce a unit antisymmetric
function
\be
\label{9}
A(x_1,\ldots,x_N) \equiv \prod_{i>j}^N {\rm sgn}(x_i-x_j) \; ,
\ee
in which
\begin{eqnarray}
{\rm sgn}(x) \equiv \frac{x}{|x|} =\left\{
\begin{array}{rc}
1, & x> 0 \; , \\
\nonumber
-1, & x< 0 \; . \end{array} \right.
\end{eqnarray}
Then the bosonic solution to Eq. (4) is given by the mapping
\be
\label{10}
\psi_B(x_1,\ldots,x_N,t) = A(x_1,\ldots,x_N)\; \psi_F(x_1,\ldots,x_N,t) \; .
\ee

By this construction, function (10) satisfies the same hard-core constraint (6)
as $\psi_F$. In the case of point particles, condition (8) holds automatically
owing  to the Pauli principle for $\psi_F$. Function (10) is totally symmetric
under permutations of any $x_i$ and $x_j$. It satisfies the initial and boundary
conditions directly following from those for $\psi_F$. In the case of a 
stationary uniform system with periodic boundary conditions, the latter are
preserved under mapping (10) if $N$ is odd, but if $N$ is even, periodic
(antiperiodic) boundary conditions on $\psi_B$ require antiperiodic (periodic)
boundary conditions on $\psi_F$. However for large $N\gg 1$, the character of
such boundary conditions becomes not important. The fermionic wave function
$\psi_F$ can be considered as corresponding to a fictitious system of spinless
fermions, or better to say, to a system of real fermions with frozen spins
aligned in the same direction. 

The Fermi-Bose mapping (10) is valid for the time-dependent wave functions,
since the antisymmetric function (9) does not include time. The system
Hamiltonian may contain any external fields and any other finite particle
interactions in addition to the hard-core ones. For the stationary 
Schr\"odinger equation (5), the mapping applies for the whole spectrum of
all eigenstates.

\section{Ground state}

It is solely for the stationary ground state that mapping (10) reduces to a
simplified form
\be
\label{11}
\psi_0^B(x_1,\ldots,x_N) = |\psi_0^F(x_1,\ldots,x_N)| \; .
\ee
As is evident, the absolute-value mapping (11) cannot apply to excited states
or to the time-dependent case, when the wave functions are, generally,
complex, whereas mapping (11) yields only real functions. In addition, for 
excited states (11) introduces unphysical cusps in $\psi_0^B$ arising from
the requirement of orthogonality of different fermionic eigenstates $\psi_0^F$,
and positivity of (11) violates the requirement of orthogonality of different
bosonic eigenstates $\psi_0^B$. These defects are not present in the 
original mapping (10).

Under the assumption that the only two-particle interaction is a zero-range
hard-core repulsion, represented by the hard-core constraint (8), and there 
are no external potentials, the ground state can be found explicitly [24]. 
Since the fermionic wave functions vanish automatically whenever any $x_i=x_j$
for $i\neq j$, the constraint has no effect, and the corresponding fermionic
ground state is that of the ideal gas of fermions, given by a Slater determinant
of the lowest $N$ single-particle plane-wave orbitals. The exact bosonic ground
state was found [24] to be a product
\be
\label{12}
\psi_0^B(x_1,\ldots,x_N) =\left [ \frac{2^{N(N-1)}}{N!L^N} \right ]^{1/2}\;
\prod_{i > j}^N \; \left | \sin\; \frac{k_0(x_i-x_j)}{N} \right | \; ,
\ee
in which $k_0$ plays the role of the Fermi wave vector,
\be
\label{13}
k_0 \equiv \pi\rho \qquad \left ( \rho\equiv \frac{N}{L} \right ) \; .
\ee

The ground-state energy can also be determined exactly [24,33,34]. For large
$N\gg 1$, it can be easily obtained from the expression
\be
\label{14}
E_0 = L \int_{-k_0}^{k_0} \; \left ( \frac{\hbar^2k^2}{2m} \right )\;
\frac{dk}{2\pi} = \frac{\hbar^2k_0^3}{6\pi m}\; L \; ,
\ee
with $k_0$ defined by the integral
\be
\label{15}
N = L \int_{-k_0}^{k_0} \; \frac{dk}{2\pi} \; ,
\ee
which yields $k_0$ from Eq. (13). Then the ground-state energy is
\be
\label{16}
E_0 = \frac{(\pi\hbar\rho)^2}{6m}\; N \; .
\ee
The lowest excitations above the ground state have a phonon character [24] 
with the sound velocity
\be
\label{17}
c =\frac{\hbar k_0}{m} \; .
\ee

The pair correlation function
\be
\label{18}
g(x,x') \equiv L^2 \int_0^L |\psi(x,x',x_3,\ldots,x_N)|^2 \; dx_3\ldots
dx_N \; ,
\ee
with the ground-state wave function (12), depends only on the difference
$x-x'$, so that $g(x,x')=g(x-x')$, and
\be
\label{19}
g(x) = 1  -\; \frac{\sin^2(k_0x)}{N^2\sin^2(k_0x/N)} \; .
\ee
For $x\ll L$, one finds [24] that
\be
\label{20}
g(x) \cong 1 - \; \frac{\sin^2(k_0x)}{(k_0x)^2} \; .
\ee
The vanishing of $g(0)=0$ at $x=0$ reflects the hard-core nature of the
two-particle interactions.

\section{Trapped gas}

The general mapping (10) holds true in the presence of any external potentials.
The case of the harmonically trapped TG gas has been considered
and an exact solution for the ground state has been obtained [31,35,36]. 
One-dimensional harmonic trap is described by the potential
\be
\label{21}
U = \frac{1}{2}\; m\om^2\; \sum_{i=1}^N x_i^2 \; .
\ee
The ground state of the Bose gas is given by mapping (11). The fermionic
ground state is a Slater determinant of the lowest $N$ single-particle
eigenfunctions $\vp_n$ of the harmonic oscillator,
$$
\vp_n(x) = \frac{\exp(-x^2/2l_0^2)}{[\sqrt{\pi}\; 2^n\;n!\;l_0]^{1/2}} \;
H_n\left ( \frac{x}{l_0}\right ) \; ,
$$
where $l_0\equiv\sqrt{\hbar/m\om}$ is the oscillator length and $H_n(\cdot)$
is a Hermite polynomial. By rearranging the corresponding fermionic determinant,
one gets [35] the Bose function
\be
\label{22}
\psi_0^B(x_1,\ldots,x_N) =  C_N \left ( \prod_{i<j}^N |x_i - x_j|
\right )\;
\prod_{i=1}^N \exp\left ( - \; \frac{x_i^2}{2l_0^2}\right ) \; ,
\ee
in which the normalization constant is
$$
C_N = \left [\frac{2^{N(N-1)/2}}{\pi^N N! (\prod_{n=0}^{N-1} n! )
l_0^N} \right ]^{1/2} \; .
$$

The single-particle density, normalized to $N$, is
\be
\label{23}
\rho(x) \equiv N \int |\psi(x,x_2,\ldots,x_N)|^2\; dx_2 \ldots dx_N \; ,
\ee
which for the ground state (22) gives
\be
\label{24}
\rho(x) = \sum_{n=0}^{N-1} |\vp_n(x)|^2 \; .
\ee
The pair correlation function becomes
\be
\label{25}
g(x,x') = 1 -\; \frac{|\Dlt(x,x')|^2}{\rho(x)\rho(x')} \; ,
\ee
where
$$
\Dlt(x,x') \equiv \sum_{n=0}^{N-1} \vp_n^*(x)\vp_n(x') \; .
$$
For large $N\gg 1$, one has $\Dlt(x,x')\approx\dlt(x-x')$. But for $x=x'$, as
is evident,
$$
\Dlt(x,x) =\rho(x) \; ,
$$
because of which function (25) vanishes, $g(x,x)=0$, as it must be for
impenetrable particles.

\section{Type of order}

What type of order exists in the system is characterized by the behaviour of
reduced density matrices [37--39]. Of particular importance is the first-order
density matrix
\be
\label{26}
\rho_1(x,x') \equiv N \int \psi(x,x_2,\ldots,x_N)\psi^*(x',x_2,\ldots,x_N)\;
dx_2 \ldots dx_N \; ,
\ee
normalized as
\be
\label{27}
\int \rho_1(x,x)\; dx = N \; .
\ee

The presence or absence of long-range order is described by the properties of
the eigenvalues $n_j$ of the density matrix (26), which are given by the
equation
\be
\label{28}
\int \rho_1(x,x') \vp_j(x')\; dx' = n_j\vp_j(x) \; .
\ee
The corresponding eigenfunctions $\vp_j(x)$ are called the natural orbitals 
[39], since in their terms the single-particle density matrix acquires a
diagonal representation
\be
\label{29}
\rho_1(x,x') = \sum_j n_j \vp_j(x) \vp_j^*(x') \; .
\ee
The eigenvalues $n_j$ play the role of the occupation numbers of the related
orbitals and are normalized as
$$
\sum_j n_j = N \; ,
$$
which results from normalization (27). The Fourier transform of matrix (26)
gives the momentum distribution
\be
\label{30}
n(k) = \int \rho_1(x,x')\; e^{-ik(x-x')}\; dx dx' \; ,
\ee
with the normalization
$$
\int n(k)\; \frac{dk}{2\pi} =  N \; .
$$

The problem of calculating the first-order density matrix (26) for a 
uniform gas was first considered by Schultz [40], who found it in the 
form of a Toeplitz determinant. Using the known asymptotics of the Toeplitz 
determinants, it was possible to prove the absence of Bose-Einstein condensate 
by showing the power-law decay of the density matrix at large distance, that 
is, by demonstrating the absence of long-range order. The precise form of 
this power-law decay was found later by Lenard [41,42], who obtained the 
long-distance behaviour as
\be
\label{31}
\rho_1(x,0) \simeq C\; \frac{\rho}{\sqrt{k_0 x}} \qquad
(|x| \ra \infty) \; ,
\ee
where $k_0\equiv\pi\rho$. The coefficient $C=0.92418$ was found by Vaidya and
Tracy [43,44]. Higher-order terms in the asymptotic behaviour of $\rho_1(x,0)$
were derived by Jimbo et al. [45]. The most accurate results are due to the
recent work by Gangardt [46], who obtained
$$
\rho_1(x,0) \simeq \frac{C\rho}{\sqrt{k_0 x}} \; \left [ 1 -\;
\frac{1}{32(k_0x)^2}\; - \; \frac{\cos(2k_0x)}{8(k_0x)^2} \; - \;
\frac{3\sin(2k_0x)}{16(k_0x)^3}\;  + \right.
$$
\be
\label{32}
\left. + \;
\frac{33}{2048(k_0x)^4} \; + \;
\frac{93\cos(2k_0x)}{256(k_0x)^4} \right ] \; .
\ee
He also found the finite-size corrections for atoms in the harmonic trapping
potential and for the case of circular geometry [46].

The investigation of the properties of the first-order density matrix (26)
revealed that the number of particles in the Bose condensate, which is
associated with the largest eigenvalue of eigenproblem (28), is of the order
of $N_0\equiv\sup_j n_j\sim\sqrt{N}$. This implies that there is no real
Bose-Einstein condensate in the uniform Tonks-Girardeau gas.

The case of trapped atoms does not allow for a simple analytical expression of
the largest eigenvalue $n_j$. The multidimensional integral (26) was evaluated
numerically by Monte Carlos integration [35,36]. Highly accurate results for 
large values of $N$ were found by Forrester et al. [47]. These results show that
again, as in the spatially uniform case, $N_0\sim\sqrt{N}$. Thus, there is no
true Bose-Einstein condensate in the trapped TG gas, because of
which the Gross-Pitaevskii equation, presuming the existence of a well-defined
order parameter associated with genuine Bose-Einstein condensate, has limited
utility, especially, for temporal processes [26,31,48]. This is contrary to 
the case of the trapped {\it ideal} gas, where Bose-Einstein condensation can
develop, though as a gradual crossover but not as a sharp phase transition 
[49]. Nevertheless, since for $N\gg 1$ the momentum distribution of the
Tonks-Girardeau gas exhibits a peak $n(k)\sim k^{-1/2}$ in the neighbourhood
of zero momentum, and a kind of order does exist, such a system displays some
coherence effects, and the Gross-Pitaevskii equation has a limited region of 
applicability [50].

In order to accurately classify the type of order arising in the
TG gas, it is possible to resort to the notion of the {\it 
order indices}, introduced for density matrices [39] and generalized for the 
case of arbitrary operators [51]. The order index of an operator $\hat A$ is 
defined [51] as
\be
\label{33}
\om(\hat A) \equiv \frac{\log||\hat A||}{\log|{\rm Tr}\hat A|} \; ,
\ee
where $||\cdot||$ means a Hermitian norm. For an $n$-th order boson density 
matrix $\hat\rho_n$, one has $||\hat\rho_n||\sim||\hat\rho_1||^n$. Also, 
$\log|{\rm Tr}\hat\rho_n|\simeq n\log N$, when $N\gg 1$. Therefore, applying 
definition (33) for a reduced density matrix $\hat\rho_n$, we have
\be
\label{34}
\om(\hat\rho_n) = \frac{\log||\hat\rho_1||}{\log N} \; .
\ee
Taking into account that
$$
||\hat\rho_1|| = \max_j n_j = N_0\; ,
$$
Eq. (34) can be rewritten as
$$
\om(\hat\rho_n) =\frac{\log N_0}{\log N} \; .
$$
For the Tonks-Girardeau gas, both uniform as well as trapped, $N_0\sim\sqrt{N}$.
Thence
\be
\label{35}
\om(\hat\rho_n) = \frac{1}{2} \; .
\ee
In the case of a genuine Bose-Einstein condensate with long-range order, one
would have $\om(\hat\rho_n)=1$. The order index (35) characterizes a system
with {\it mid-range} order [39,51]. The occurrence of mid-range order means that,
though there is no true Bose-Einstein condensate, some partial coherence
does exist in the system.

\section{Lieb-Liniger gas}

Up to now, the Fermi-Bose mapping (10) has been applied to the TG
gas, that is, the one-dimensional gas of impenetrable bosons. However, as it
turned out, the applicability of this mapping is much wider, being valid for
a large class of one-dimensional systems called the Lieb-Liniger gas, which
is characterized by the contact two-particle interaction
\be
\label{36}
\Phi(x) = \Phi_0\dlt(x) \; .
\ee

The one-dimensional system with this interaction was studied by Lieb and Liniger
[52,53]. The delta potential (36) leaves the wave function continuous but yields
a jump in the derivative according to the condition
\be
\label{37}
\left ( \frac{\prt}{\prt x_i} \; - \; \frac{\prt}{\prt x_j} \right )
\psi_B{\Big |}_{x_i=x_j+0} = -\left ( \frac{\prt}{\prt x_i} \; - \;
\frac{\prt}{\prt x_j} \right )\psi_B{\Big |}_{x_i=x_j-0} = 
\frac{m\Phi_0}{\hbar^2}\; \psi_B {\Big |}_{x_i=x_j\pm 0} \; .
\ee
The dimensionless ground-state energy of this gas,
\be
\label{38}
e(g) \equiv \frac{2mE_0}{\hbar^2\rho^2N} \; ,
\ee
expressed as a function of the dimensionless coupling parameter
\be
\label{39}
g \equiv \frac{m\Phi_0}{\rho\hbar^2}\; ,
\ee
is given [52] by the equation
\be
\label{40}
e(g) = \left ( \frac{g}{\lbd}\right )^3 \; 
\int_{-1}^1 f(x)x^2\; dx \; ,
\ee
in which the function $f(x)$ satisfies the integral equation
\be
\label{41}
2\pi f(x) = 1 + 2\lbd \int_{-1}^1  \; \frac{f(y)\; dy}{\lbd^2+(x-y)^2} \; ,
\ee
and the constant $\lbd$ is defined by the normalization condition
\be
\label{42}
\frac{g}{\lbd} \; \int_{-1}^1 \; f(x)\; dx = 1 \; .
\ee
Numerical solution for $e(g)$ was given in Refs. [52,54]. A detailed table 
can be found on the website [55]. An analytical asymptotic expansion in the 
weak-coupling limit reads as
\be
\label{43}
e(g) \simeq g + c_3 g^{3/2} + c_4 g^2 + c_5 g^{5/2} \; .
\ee
The coefficients
$$
c_3 = -\frac{4}{3\pi}=-0.424413 \; , \qquad c_4=\frac{1.29}{2\pi^2}=0.065352
$$
were found by Lee [56,57]. The coefficient $c_5$ is not known exactly. Its
estimate is $c_5=-0.017201$.  The strong-coupling expansion, being based on the
numerical results [55], can be derived as
\be
\label{44}
e(g) \simeq e(\infty) \left ( 1 -\; \frac{4}{g}\; + \;
\frac{12}{g^2}\; - \; \frac{32}{g^3}\; + \; \frac{80}{g^4}
\right ) \; ,
\ee
where
\be
\label{45}
e(\infty) \equiv \frac{\pi^2}{3}
\ee
is the TG limit [24]. The first-order density matrix and the pair 
correlation function were investigated by Monte Carlo techniques [58]. It has
been mentioned that mapping (10) can serve as a reasonable approximation for
limited time intervals of the time-dependent Schr\"odinger equation [59].

Cheon and Shigehara [60,61] showed that mapping (10) is {\it exact} for the
Lieb-Liniger gas, provided that the fermionic wave function satisfies the
condition
\be
\label{46}
\left ( \frac{\prt}{\prt x_i} \; - \; \frac{\prt}{\prt x_j} \right )
\psi_F{\Big |}_{x_i=x_j\pm 0} =
\frac{m\Phi_0}{\hbar^2}\; \psi_F {\Big |}_{x_i=x_j+0} =-\;
\frac{m\Phi_0}{\hbar^2}\; \psi_F {\Big |}_{x_i=x_j-0}\; .
\ee
Here the derivative is continuous but the wave function is discontinuous.

Recently a new development has been proposed [35,62] for the case of the 
spin-aligned Fermi gas, suggesting exploitation of the generalized Fermi-Bose 
mapping [60--64] in the {\it opposite} direction, by mapping the fermionic 
Tonks-Girardeau gas, a spin-aligned Fermi gas with strong one-dimensional 
atomic interactions mediated by a three-dimensional $p$-wave Feshbach
resonance, to the trapped ideal Bose gas.

It can be shown [62,64] that in the low-energy domain the one-dimensional 
longitudinal scattering of two spin-aligned fermions, confined in a 
single-mode harmonic waveguide, can be well represented by the contact 
condition
\be
\label{47}
\psi_F{\large |}_{x_i=x_j-0} = -\psi_F{\large |}_{x_i=x_j+0} =
a_{1D}\; \left. \frac{\prt\psi_F}{\prt x}\right |_{x_i=x_j\pm 0} \; .
\ee
Here the following notation is used for the effective one-dimensional 
scattering length:
\be
\label{48}
a_{1D} = \frac{3a_p^3}{l_\perp^2}\left [ 1 + 
\frac{3\zeta(3/2)}{2\sqrt{2}\pi} \left ( \frac{a_p}{l_\perp}\right )^3
\right ]^{-1} \; ,
\ee
in which $a_p$ is the $p$-wave scattering length, $\zeta(3/2)=2.612$ is 
the Riemann zeta function, and $l_\perp\equiv\sqrt{\hbar/m\om_\perp}$ is 
the transverse oscillator length. The scattering length $a_{1D}$ diverges 
at $(a_p/l_\perp)^3\cong -1.134$. The fermionic TG gas regime 
occurs in the neighbourhood of this resonance. The one-dimensional scattering 
lengths are invariant under the Fermi-Bose mapping (10), as a result of which 
the scattering length $a_{1D}$ is the same for bosons and fermions.

The contact condition (47) is generated by the one-dimensional pseudopotential
operator [62]
\be
\label{49}
\Phi^F(x) =  \Phi_0^F\dlt'(x)\; \frac{1}{2}\left (
\left.\frac{\prt\psi_F}{\prt x} \right |_{x+0} -\; \left.
\frac{\prt\psi_F}{\prt x}\right |_{x-0}\right ) \; ,
\ee
in which $\dlt'(x)$ is the derivative of the Dirac delta-function and 
the effective coupling strength is
$$
\Phi_0^F = 2\hbar^2 \; \frac{a_{1D}}{m} \; .
$$
This should be compared with the bosonic interaction strength
$$
\Phi_0^B = -\; \frac{2\hbar^2}{m a_{1D}} \; .
$$
The spin-aligned Fermi gas maps to the Lieb-Liniger Bose gas, with the 
fermionic and bosonic interaction strengths inversely related [60,61] as
\be
\label{50}
\Phi_0^B \Phi_0^F = -\; \frac{4\hbar^4}{m^2} \; .
\ee
From here, it is clear that a strongly interacting Fermi gas can be 
mapped to a weakly interacting Bose gas. In the limiting case, when at 
the resonance the fermion interaction becomes divergent so that 
$a_{1D}\ra-\infty$, the corresponding Bose gas is asymptotically free.

\section{Discussion}

In the present survey, we have considered one-dimensional Bose gases. The
properties of such gases are drastically different from those of their
three-dimensional counterparts. For comparison, we may recall the ground-state
energy of the three-dimensional dilute Bose gas with a hard-sphere interaction.
The dimensionless ground-state energy is
$$
\frac{2mE_0}{\hbar^2\rho^{2/3}N} \approx 4\pi\al^{1/3}\left ( 1 +
b_1 \al^{1/2} + b_2\al + b_2'\al\ln\al \right ) \; ,
$$
where $\al\equiv\rho a^3\ll 1$, with $a$ being the sphere diameter, and where
the coefficients $b_1$ and $b_2'$ are
$$
b_1  = \frac{128}{15\sqrt{\pi}} =  4.814418 \; , \qquad
b_2'=8\left ( \frac{4\pi}{3}\; - \; \sqrt{3}\right ) =
19.653915 \; .
$$
The coefficient $b_1$ was found in Refs. [65--69] and $b_2'$ in Refs. 
[70,71]. The coefficient $b_2$ has not been determined exactly. According 
to Wu [70], one has $b_2=b_2'\ln(12\pi)=71.337$. Hugenholtz and Pines [72] 
give $b_2\approx 74.617$. Another quantity that could be compared with its 
one-dimensional counterpart is the pair correlation function of the 
three-dimensional dilute gas of hard-sphere bosons, which is found [67] 
to be
$$
g(r) \simeq 1 - \; \frac{C}{\al^{1/6}} \left ( \frac{a_0}{r}\right )^4 \; ,
$$
where $r\gg a_0$, $\al\ll 1$ and
$$
C= \frac{1}{(2\pi)^2\sqrt{\pi}} \; , \qquad
\rho a_0^3 \equiv 1 \; ,
$$
$a_0$ being the mean interparticle distance. The behaviour of both $E_0$ and
$g(r)$ for the three-dimensional gas is essentially different from their
one-dimensional analogs. Being principally different, the three-dimensional
gas does not allow for a direct mapping between bosons and fermions, as in the
one-dimensional case, which makes its treatment much more involved. While in 
the one-dimensional case, the Fermi-Bose mapping (10) gives a great advantage
in the simplification of the mathematical description.

Nowadays one-dimensional Bose gases are not just artificial models, but vise
versa are real physical objects realized in a number of experiments [5--21]
with quasi-one-dimensional traps and waveguides. Quasi-one-dimensional dilute
Bose gases can be treated as purely one-dimensional gases, with their effective
interaction of the contact type (36), where the interaction strength
$$
\Phi_0 = \frac{2\hbar^2 a_s}{ml_\perp^2}\left [ 1 -\; 
\frac{|\zeta(1/2)|a_s}{\sqrt{2}l_\perp} \right ]^{-1}
$$
is expressed through the three-dimensional scattering length $a_s$ and the
oscillator length $l_\perp\equiv\sqrt{\hbar/m\om_\perp}$ of the confining
radial potential [73,74]. Here $\zeta(1/2)=-1.4603$.

An additional dimension is provided by the Feshbach resonance techniques 
which make it possible: to vary the scattering length in a very wide range,
transforming the interaction strength from weak to strong coupling; to 
realize heteronuclear resonances between two different atomic species 
[75,76]; and to perform a crossover between a degenerate fermionic gas and 
a gas of bosonic molecules (see Refs. [23,77--79]). In quasi-one-dimensional 
traps, the Feshbach resonance techniques would allow for a continuous tuning 
of the system properties from a weakly interacting gas to the hard-core
TG gas [80].

The remarkable generality of the Fermi-Bose mapping (10) provides us with a
convenient practical tool for considering different regimes of one-dimensional
or quasi-one-dimensional dilute Bose systems, from the TG gas
of impenetrable bosons to the Lieb-Liniger gas with contact interactions.
Moreover, mapping (10) is applicable for describing nonequilibrium processes in
such one-dimensional gases. A great variety of available experiments, to 
which the general mapping (10) is applicable, makes the latter of fundamental
importance.


\newpage

\begin{thebibliography}{99}

\bibitem{1}
P.W. Courteille, V.S. Bagnato, and V.I. Yukalov, Laser Phys. {\bf 11},
659 (2001).

\bibitem{2}
L. Pitaevskii and S. Stringari, Bose-Einstein Condensation (Clarendon, Oxford,
2003).

\bibitem{3}
J.O. Andersen, Rev. Mod. Phys. {\bf 76}, 599 (2004).

\bibitem{4}
K. Bongs and K. Sengstock, Rep. Prog. Phys. {\bf 67}, 907 (2004).

\bibitem{5}
J. Denschlag, D. Cassettari, and J. Schmiedmayer, Phys. Rev. Lett. {\bf 82},
2014 (1999).

\bibitem{6}
J.H. Thywissen, R.M. Westervelt, and M. Prentiss, Phys. Rev. Lett. {\bf 83},
3762 (1999).

\bibitem{7}
D. M\"uller, D.Z. Andersen, R.J. Grow, et al., Phys. Rev. Lett. {\bf 83}, 5194
(1999).

\bibitem{8}
N.H. Dekker, C.S. Lee, V. Lorent, et al., Phys. Rev. Lett. {\bf 84}, 1124 
(2000).

\bibitem{9}
M. Key, I.G. Hughes, W. Rooijakkers, et al., Phys. Rev. Lett. {\bf 84}, 1371
(2000).

\bibitem{10}
K. Bongs, S. Burger, S. Dettmer, et al., Phys. Rev. A {\bf 63}, 031602 (2001).

\bibitem{11}
J. Arlt and K. Dholakia, Phys. Rev. A {\bf 63}, 063602 (2001).

\bibitem{12}
F. Schreck, L. Khaykovich, K.L. Corwin, et al., Phys. Rev. Lett. {\bf 87},
080403 (2001).

\bibitem{13}
A. G\"orlitz, J.M. Vogels, A.E. Leanhardt, et al., Phys. Rev. Lett. {\bf 87},
130402 (2001).

\bibitem{14}
M. Greiner, I. Bloch, O. Mandel, et al., Phys. Rev. Lett. {\bf 87},
160405 (2001).

\bibitem{15}
S. Dettmer, D. Hellweg, P. Ryytty, et al., Phys. Rev. Lett. {\bf 87},
160406 (2001).

\bibitem{16}
D. Hellweg, S. Dettmer, P. Ryytty, et al., Appl. Phys. B {\bf 73}, 781 (2001).

\bibitem{17}
S. Richard, F. Gerbier, J.H. Thywissen, et al., Phys. Rev. Lett. {\bf 91},
010405 (2003).

\bibitem{18}
F. Gerbier, J.H. Thywissen, S. Richard, et al., Phys. Rev. A {\bf 67},
051602 (2003).

\bibitem{19}
H. Moritz, T. St\"oferle, M. K\"ohl, and T. Esslinger, Phys. Rev. Lett.
{\bf 91}, 250402 (2003).

\bibitem{20}
B. Paredes, A. Widera, V. Murg, et al., Nature {\bf 429}, 277 (2004).

\bibitem{21}
T. Kinoshita, T. Wenger, and D.S. Weiss, Science {\bf 305}, 1125 (2004).

\bibitem{22}
D.S. Petrov, D.M. Gangardt, and G.V. Shlyapnikov, J. Physique {\bf 116}, 3
(2004).

\bibitem{23}
V.I. Yukalov, Laser Phys. Lett. {\bf 1}, 435 (2004).

\bibitem{24}
M. Girardeau, J. Math. Phys. {\bf 1}, 516 (1960).

\bibitem{25}
M.D. Girardeau, Phys. Rev. B {\bf 139}, 500 (1965).

\bibitem{26}
M.D. Girardeau and E.M. Wright, Phys. Rev. Lett. {\bf 84}, 5239 (2000).

\bibitem{27}
M.D. Girardeau and E.M. Wright, Phys. Rev. Lett. {\bf 84}, 5691 (2000).

\bibitem{28}
K.K. Das, G.J. Lapeyre, and E.M. Wright, Phys. Rev. A {\bf 65}, 063603 (2002).

\bibitem{29}
M.D. Girardeau, K.K. Das, and E.M. Wright, Phys. Rev. A {\bf 66}, 023604 (2002).

\bibitem{30}
K.K. Das, M.D. Girardeau, and E.M. Wright, Phys. Rev. Lett. {\bf 89}, 170404 
(2002).

\bibitem{31}
 M.D. Girardeau and E.M. Wright, Laser Phys. {\bf 12}, 8 (2002).

\bibitem{32}
L. Tonks, Phys. Rev. {\bf 50}, 955 (1936).

\bibitem{33}
A. Bijl, Physica {\bf 4}, 329 (1937).

\bibitem{34}
T. Nagamiya, Proc. Phys. Math. Soc. Japan, {\bf 22}, 705 (1940).

\bibitem{35}
M.D. Girardeau, E.M. Wright, and J.M. Triscari, Phys. Rev. A {\bf 63}, 033601
(2001).

\bibitem{36}
G.J. Lapeyre, M.D. Girardeau, and E.M. Wright, Phys. Rev. A {\bf 66}, 023606
(2002).

\bibitem{37}
O. Penrose and L. Onsager, Phys. Rev. {\bf 104}, 576 (1956).

\bibitem{38}
C.N. Yang, Rev. Mod. Phys. {\bf 34}, 694 (1962).

\bibitem{39}
A.J. Coleman and V.I. Yukalov, Reduced Density Matrices (Springer, Berlin,
2000).

\bibitem{40}
T.D. Schultz, J. Math. Phys. {\bf 4}, 666 (1963).

\bibitem{41}
A. Lenard, J. Math. Phys. {\bf 5}, 930 (1964).

\bibitem{42}
A. Lenard, J. Math. Phys. {\bf 7}, 1268 (1966).

\bibitem{43}
H.C. Vaidya and C.A. Tracy, Phys. Rev. Lett. {\bf 42}, 3 (1979).

\bibitem{44}
H.C. Vaidya and C.A. Tracy, J. Math. Phys. {\bf 20}, 2291 (1979).

\bibitem{45}
M. Jimbo, T. Miwa, Y. Mori, and M. Sato, Physica D {\bf 1}, 80 (1980).

\bibitem{46}
D.M. Gangardt, e-print cond-mat/0404104 (2004).

\bibitem{47}
P.J. Forrester, N.E. Frankel, T.M. Garoni, and N.S. Witte, Phys. Rev. A
{\bf 67}, 043607 (2003).

\bibitem{48}
M.D. Girardeau and E.M. Wright, e-print cond-mat/0010457 (2000).

\bibitem{49}
W. Ketterle and N.J. Van Druten, Phys. Rev. A {\bf 54}, 656 (1996).

\bibitem{50}
E.B. Kolemeisky, T.J. Newman, J.P. Straley, and X. Qi, Phys. Rev. Lett. 
{\bf 85}, 1146 (2000).

\bibitem{51}
V.I. Yukalov, Physica A {\bf 310}, 413 (2002).

\bibitem{52}
E.H. Lieb and W. Liniger, Phys. Rev. {\bf 130}, 1605 (1963).

\bibitem{53}
E.H. Lieb, Phys. Rev. {\bf 130}, 1616 (1963).

\bibitem{54}
V. Dunjko, V. Lorent, and M. Olshanii, Phys. Rev. Lett. {\bf 86}, 5413 (2001).

\bibitem{55}
V. Dunjko and M. Olshanii, http://physics.usc.edu/$\sim$olshanii/DIST/.

\bibitem{56}
D.K. Lee, Phys. Lett. A {\bf 37}, 49 (1971).

\bibitem{57}
D.K. Lee, Phys. Rev. A {\bf 3}, 345 (1971).

\bibitem{58}
G.E. Astrakharchik and S. Giorgini, Phys. Rev. A {\bf 68}, 031602 (2003).

\bibitem{59}
M.D. Girardeau, Phys. Rev. Lett. {\bf 91}, 040401 (2003).

\bibitem{60}
T. Cheon and T. Shigehara, Phys. Lett. A {\bf 243}, 111 (1998).


\bibitem{61}
T. Cheon and T. Shigehara, Phys. Rev. Lett. {\bf 82}, 2536 (1999).

\bibitem{62}
M.D. Girardeau, H. Nguyen, and M. Olshanii, Opt. Commun. {\bf 243}, 3 (2004).

\bibitem{63}
M.D. Girardeau, and M. Olshanii, e-print cond-mat/0309396 (2003).

\bibitem{64}
B.E. Granger and D. Blume, Phys. Rev. Lett. {\bf 92}, 133202 (2004).

\bibitem{65}
T.D. Lee and C.N. Yang, Phys. Rev. {\bf 105}, 1119 (1957).

\bibitem{66}
T.D. Lee, K. Huang, and C.N. Yang, Phys. Rev. {\bf 106}, 1135 (1957).

\bibitem{67}
T.D. Lee and C.N. Yang, Phys. Rev. {\bf 112}, 1419 (1958).

\bibitem{68}
S.T. Beliaev, J. Exp. Theor. Phys. {\bf 7}, 289 (1958).

\bibitem{69}
S.T. Beliaev, J. Exp. Theor. Phys. {\bf 7}, 299 (1958).

\bibitem{70}
T.T. Wu, Phys. Rev. {\bf 115}, 1390 (1959).

\bibitem{71}
K. Sawada, Phys. Rev. {\bf 116}, 1344 (1959).

\bibitem{72}
N.M. Hugenholtz and D. Pines, Phys. Rev. {\bf 116}, 489 (1959).

\bibitem{73}
M. Olshanii, Phys. Rev. Lett. {\bf 81}, 938 (1998).

\bibitem{74}
D.S. Petrov, G.V. Schlyapnikov, and J.T.M. Walraven, Phys. Rev. Lett. 
{\bf 85}, 3745 (2000).

\bibitem{75}
C.A. Stan, M.W. Zwierlein, C.H. Schunk, et al., e-print cond-mat/0406129
(2004).

\bibitem{76}
S. Inouye, J. Goldwin, M.L. Olsen, et al., e-print cond-mat/0406208 (2004).

\bibitem{77}
S.J.J.M.F. Kokkelmans, G.V. Shlyapnikov, and S. Salomon, Phys. Rev. A
{\bf 69}, 031602 (2004).

\bibitem{78}
R.A. Duine and H.T.C. Stoof, Phys. Rep. {\bf 396}, 115 (2004).

\bibitem{79}
Q. Chen, J. Stajic, S. Tan, and K. Levin, e-print cond-mat/0404274 
(2004).

\bibitem{80}
I.V. Tokatly, Phys. Rev. Lett. {\bf 93}, 090405 (2004).

\end{thebibliography}
\end{document}